\documentclass[12pt]{article}

\newcommand{\add}{\addtocounter{eqncnt}{1}}
\newcounter{eqncnt}[section]

\newcommand{\be}{\begin{equation}}
\newcommand{\ee}{\end{equation}\add}
\newcommand{\bea}{\begin{eqnarray}}
\newcommand{\eea}{\end{eqnarray}}

\newcommand{\LCDM}{$\Lambda$CDM\ }

\begin{document}
\begin{center}
{\Large \bf Cosmological Observations: Averaging on the Null Cone} \\[2mm]

\vskip .5in

{\sc A.A. Coley}\\
{\it Department of Mathematics and Statistics}\\
{\it Dalhousie University, Halifax, NS B3H 3J5, Canada}\\
{\it aac@mathstat.dal.ca }
\end{center}

\begin{abstract}

The universe is not isotropic or spatially homogeneous on local
scales. The averaging of local inhomogeneities in general
relativity can lead to significant dynamical effects on the
evolution of the universe and on the interpretation of
cosmological data. In particular, all deductions about cosmology
are based on light paths; averaging can have an important effect
on photon propagation and hence cosmological observations. It
would be desirable to describe the physical effects of averaging
in terms of observational quantities and focussing on the
behaviour of light. Data (e.g., matter terms, such as the density
or galaxy number counts, which are already expressed as averaged
quantities) is given on the null cone. Therefore, it is
observationally meaningful to consider light-cone averages of
quantities. In principle, we wish to describe the cosmological
equations on the null cone, and hence we need to construct the
averaged geometry on the null cone. However, we argue that it is
still necessary to average the full Einstein field equations to
obtain suitably averaged equations on the null cone. Since it is
not the geometry per se that appears in the observational
relations, we discuss whether it is possible to covariantly
`average' just a subset of the evolution equations on the null
cone, focussing on relevant observational quantities. We present
an averaged version of the scalar null Raychaudhuri equation,
which may be a useful first step in this regard.

\end{abstract}

\newpage


\newpage

\section{Introduction}

Cosmological observations \cite{Riess:2006fw,bennett}, based on
the assumption of a spatially homogeneous and isotropic
Friedmann-Lema\^{i}tre-Robertson-Walker (FLRW) model plus small
perturbations, are usually interpreted as implying that there
exists dark energy, the spatial geometry is flat, and that there
is currently an accelerated expansion giving rise to the so-called
$\Lambda$CDM-concordance model. Although the concordance model is
quite remarkable (at least if the idea of dark energy can be
tolerated), it does not convincingly fit all data. Essentially,
structure is more evolved on large scales than predicted by the
$\Lambda$CDM-concordance model. There is strong evidence for
coherent bulk flows on scales out to at least 300$h^{-1}$Mpc
\cite{Kash} and there is observational evidence of anisotropy in
the Hubble expansion rate, suggesting large scale peculiar motions
\cite{probs3}. In addition, there are correlations between galaxy
surveys and the cosmic microwave background (CMB) data
\cite{probs1}, overestimates of the amplitude of the matter power
spectrum \cite{probs2} and non-Gaussianities in the CMB
\cite{probs4}. Unfortunately, if the underlying cosmological model
is not a perturbation of an exact flat FLRW solution, the
conventional data analysis and their interpretation is not
necessarily valid. For example, the standard analysis of type Ia
supernovae (SNIa) and CMB data in FLRW models cannot be applied
directly when backreaction effects are present, because of the
different behaviour of the spatial curvature \cite{Shapiro:2005}.

Supernovae data {\it dynamically} requires an accelerating
universe. However, this only implies the existence of dark energy
if the universe  is (approximately) FLRW. Thus the isotropic and
spatially homogeneous $\Lambda$CDM model is a good {\it
phenomenological} fit to the real inhomogeneous universe, as far
as observational determinations of the expansion history of the
universe \cite{marra-sc}. However, this does not imply that a
primary source of dark energy exists, but only that it exists as
far as the phenomenological fit is concerned. Supernovae data can
be explained without dark energy in inhomogeneous models, where
the full effects of general relativity (GR) come into play. The
apparent acceleration of the universe is thus not caused by
repulsive gravity due to dark energy, but rather is a dynamical
result of inhomogeneities, either in an exact solution or via
averaging effects (due to the back-reaction of inhomogeneities).
Indeed, it has been indeed shown that the
Lema\^{i}tre-Tolman-Bondi (LTB) solution can be used to fit the
observed data without the need of dark energy \cite{LTBgeo},
although it is necessary to place the observer at the center of a
rather large-scale underdensity.

Therefore, the averaging problem in cosmology is of considerable
importance for the correct interpretation of cosmological data.
The correct governing equations on cosmological scales are
obtained by averaging the Einstein field equations (EFE) of GR
(plus a theory of photon propagation; i.e., information on what
trajectories actual particles follow). By assuming spatial
homogeneity and isotropy on the largest scales, the
inhomogeneities affect the dynamics though correction
(backreaction) terms, which can lead to behaviour qualitatively
and quantitatively different from the FLRW models. It is necessary
to use an exact covariant approach which gives a prescription for
the correlation functions that emerge in an averaging of the full
tensorial EFE. For example, in \cite{CPZ} the macroscopic gravity
equations were explicitly solved in a FLRW background geometry and
it was found that the correlation tensor (backreaction) is of the
form of a spatial curvature.

Clearly, backreaction (averaging) effects are real, but their
relative importance must be determined.  In the  FLRW plus
perturbations approach, the (backreaction) effects are assumed
small (and are {\em assumed} to stay small during the evolution of
the universe). But it is also possible that averaging effects are
not small (i.e., perturbation theory cannot be used to estimate
the effects, and real inhomogeneous effects must be included). The
Wilkinson Microwave Anisotropy Probe (WMAP)  \cite{bennett},
together with  SNIa data in $\Lambda$CDM models
\cite{Riess:2006fw}, suggests a normalized spatial curvature
$\Omega_{k} \approx 0.01 - 0.02$ (i.e., of about a percent).
Combining these observations with large scale structure
observations then puts stringent limits on the curvature parameter
in the context of adiabatic $\Lambda$CDM models; however, these
data analyses are very model- and prior-dependent
\cite{Shapiro:2005}, and care is needed in the proper
interpretation of the data. There is a heuristic argument that
$\Omega_{k}\sim 10^{-3}-10^{-2}$ \cite{NewRas,Col}, which is
consistent with CMB observations \cite{Col} and agrees with
estimates for intrinsic curvature fluctuations using realistically
modelled clusters and voids in a Swiss-cheese model
\cite{buch,hellaby:volumematching}. It must be appreciated that
such a value for $\Omega_{k}$, at the 1\% level, is relatively
large and may have a significant dynamical effect on the evolution
of the universe; the correction terms change the interpretation of
observations so that they need to be accounted for carefully to
determine if a model may be consistent with cosmological data
\cite{dunkley,Col}.

\section{Null geodesics}

All deductions about cosmology are based on light paths. Only the
redshift and the energy flux of light arriving from a distant
source are observed, rather than the expansion rate or the matter
density. It is often assumed that intervening inhomogeneities
average out. However, inhomogeneities affect curved null geodesics
\cite{NewRas,buch} and can drastically alter observed distances
when they are a sizable fraction of the curvature radius. In the
real universe, voids occupy a much larger region as compared to
structures \cite{Hoyle:2003}, hence light preferentially travels
much more through underdense regions and the effects of
inhomogeneities on luminosity distance are likely to be
significant.

The effect of averaging null geodesics in inhomogeneous models was
discussed in \cite{coleynull}.  GR is treated as a microscopic
(classical) theory. Real photons travel on null geodesics in the
microscopic geometry. However, because all observations are of
finite resolution, observations necessarily involve averages of
measured quantities. Therefore, in interpreting real observations,
it is necessary to model properties of (not only a single photon
but of) a `narrow' beam or bundle of photons (i.e., a local
congruence of null geodesics). From the geometric optics
approximation we can obtain the optical scalar (Dyer-Roeder)
equations that govern the propagation of the local shearing and
expansion (of the cross-sectional area of the beam) with respect
to the affine parameter along the congruence due to Ricci
focussing and Weyl tidal focussing \cite{Dyer}. Since the {\em
nonlinear} optical scalar equations require integration along the
beam, the optics for a lumpy distribution does not average and
there may be important resulting effects. Therefore, it is
important to study the effect of averaging on a beam of photons in
the optical limit.

The motion of photons in an averaged geometry (and the resulting
effect on cosmological observations) was discussed in
\cite{coleynull}.  Assuming GR to be  a microscopic theory on
small scales, with local metric field ${\bf g}$ (the
micro-geometry) and matter fields, a photon follows a null
geodesic ${\bf k}$ in the local geometry.  After averaging (using
a covariant averaging scheme), we obtain a smoothed out
macroscopic geometry (with macroscopic metric $\langle {\bf g}
\rangle$) and macroscopic matter fields, valid on larger scales.
But what trajectories do photons follow in the macro-geometry?  In
general, the ``averaged'' vector $\langle {\bf k} \rangle$ need
not be null,  need not be geodesic (and even if it is, need not be
affinely parametrized) in the macro-geometry. This clearly affects
cosmological observations. In the case of radial geodesics in a
sphericallly symmetric geometry \cite{coleynull}, it was found
that there is an effect due to the non-affine parameterization of
the null geodesics in the averaged geometry, together with
additional small corrections due to the fact that the trajectories
are not exactly null geodesics, the correlation tensor is not
precisely due to a spatial curvature and since a beam of photons
will experience expansion and shearing during its evolution
\cite{Dyer}. Each of these effects are typically of order of about
1\%, which can add up and perhaps produce a significant
observational effect.

Similar issues have been discussed recently from a different point
of view. It was conjectured that in a statistically homogeneous
and isotropic dust universe, light propagation can be treated in
terms of the overall geometry (meaning the average expansion rate
and average spatial curvature) if the structures are realistically
small and the observer is not in a special location
\cite{{NewRas}}. The relationship between the expansion rate, the
redshift and the distance scale in such a universe containing
non-linear structures was further investigated in  \cite{R2009},
and it was shown that light propagation can be expressed in terms
of averaged geometrical quantities, up to a term related to the
null geodesic shear. In general, the null shear is not negligible,
and thus the Dyer-Roeder equations do not correctly describe the
effect of clumping. Instead, the redshift and the distance are
determined by the average expansion rate, the matter density today
and the null geodesic shear. This implies that a clumpy model can
be consistent with the observed position of the CMB acoustic peaks
even when there is significant spatial curvature \cite{bennett},
provided that the expansion history is sufficiently close to the
spatially flat \LCDM model \cite{{R2009}}.

In addition, the propagation of photons in a particular (toy)
Swiss-cheese model, where the cheese consists of a spatially flat,
matter only FLRW solution and the holes are constructed out of a
LTB solution of the EFE, and the phenomenological effects of
large-scale nonlinear inhomogeneities on observables such as the
luminosity-distance--redshift relation, were discussed in
\cite{marra-sc}. Following a fitting procedure based on light-cone
averages, it was found that the expansion scalar is unaffected by
the inhomogeneities (essentially due to the spherical symmetry of
the model), but the light-cone average of the density as a
function of redshift (which is not affected by the spherical
symmetry) is affected by inhomogeneities.  (The effect arises
because, as the universe evolves,  a photon spends more and more
time in the (large) voids than in the (thin) high-density
structures.)  The phenomenological homogeneous model describing
the light-cone average of the density is similar to the
$\Lambda$CDM concordance model, and behaves as if it has a
dark-energy component.

Clearly averaging can have an important effect on photon
propagation and hence observations. Ideally, we would like to
describe the physical effects of averaging in terms of
observational quantities, focussing on the behaviour of light. In
particular, spatial averages at constant time are not directly
related  to observations; it is observationally more meaningful to
consider light-cone averages of quantities. Consequently, we wish
to attempt to describe the evolution equations on the null cone,
focussing on observational quantities. Let us first consider the
formulation of the EFE on the null cone.

\section{EFEs in Observational Coordinates}

The cosmological data representing galaxy redshifts, observer area
distances and galaxy number counts as functions of redshift are
given, not on a space-like surface of constant time, but rather on
our past light cone $C^-(p_0)$, which is centered at our
observational position $p_0$ ``here and now'' on our world line $
{\cal C}$.  By using {\em observational coordinates} (OC)
\cite{Ellis et al},  the EFE can be formulated in a way which
reflects both the geodesic flow of the cosmological fluid and the
null geometry of $C^-(p_0)$, along which all of the observational
information reaches us. In this formulation the EFE split
naturally into a set of equations which can be solved on
$C^-(p_0)$, that is on our past light cone, and a second set which
evolves these solutions off $C^-(p_0)$ to other light cones into
the past or into the future. The solution to the first set is
directly determined from the data, and those solutions constitute
the ``initial conditions'' for the solution of the second set.

For example, the OC $x^i=\{w,y,\theta ,\phi \}$ for a spherically
symmetric metric are centered on the observer's world line $C$ and
defined in the following way: (i) $w$ is constant on each past
light cone along $C$; each observational time coordinate $w =
constant$ specifies a past light cone along $C$ and our past light
cone is designated as $w = w_0$. (ii) $y$ is the null radial
coordinate, which measures distance down the null geodesics with
affine parameter $v$, generating each past light cone centered on
$C$, so that $y$ increases as one moves down a past light cone
away from $C$. (iii) $\theta$ and $\phi$ are usual the latitude
and longitude of observation, respectively. In OC the spherically
symmetric metric takes the form:
\begin{equation}
ds^2=-A(w,y)^2dw^2+2A(w,y)B(w,y)dwdy+C(w,y)^2d\Omega ^2,
\label{oc}
\end{equation}
where the comoving fluid 4-velocity is $u^a=A^{-1}\delta _w^a$.
There is remaining coordinate freedom which preserves the
observational form of the metric.
It is also important to specify the central conditions for the metric variables $%
A(w, y)$, $B(w, y)$ and $C(w, y)$ in equation (\ref{oc})~(as they
approach $y = 0$). The FLRW metric is obtained in OC by
effectively specifying $A=B=a(\eta), C= a(\eta) k^{-2} sin^2(ky)$
in conformal time $\eta = w - y$. It was shown in \cite{AS} how to
construct flat dust-filled $\Lambda \neq 0$ FLRW cosmological
models from FLRW cosmological data on our past light cone, by
integrating the exact spherically symmetric EFE in OC first on the
light cone (integrating, e.g., ${\frac{C^{\prime\prime}}{C}}$) and
then off of it  (integrating, e.g., ${\frac{{\ddot{C}}}C}$).

The basic observational quantities on $C$ are then defined as follows:

(i) The redshift $z$ at time $w_0$ on $C$ for a comoving
source a null radial distance $y$ down $C^{-}(p_0)$ is given by
$1+z=  {A(w_0,0)}/{A(w_0,y)}$.

(ii) The luminosity distance $d_L$ \cite{Ellis et al}, measured at
time $w_0$ on $C$ for a source at a null radial distance $y$, is
given by $d_L = (1+z)^2 C(w_0, y)$.

(iv) The number of galaxies counted by a central observer out to a
null radial distance $y$ is given by $N(y)=4\pi\int_0^y
\mu(w_0,\tilde{y})m^{-1}B(w_0,\tilde{y})C(w_0,\tilde{y})^2
d\tilde{y}$, where $\mu$ is the mass-energy density and $m$ is the
average galaxy mass.

Data is given on the null cone (in a suitably averaged form). In
principle, we need to construct the metric on the null cone. For
example, in the spherically symmetric case averaged quantities are
given as simple integrations with respect to y (or redshift);
matter terms, such as $N(y)$, are already expressed as averaged
quantities. To obtain the averaged geometry on the null cone
(i.e., in terms of averaged metric functions $\langle {A}
\rangle$, $\langle {B} \rangle$ and $\langle {C} \rangle$), we
still need to average the EFE on the null cone. Thus we still need
to be able to average the tensorial EFE using some covariant
averaging scheme, and then separate the equations into evolution
equations on the null cone (in terms of $\langle {A} \rangle$,
$\langle {B} \rangle$ and $\langle {C} \rangle$ and their
derivatives), and constraints and evolution equations off the null
cone. Therefore, we need to project the suitably covariantly
averaged full EFE onto the null cone to compare with data averaged
on the null cone.

Thus, in this approach we are no further ahead, unless we can
covariantly `average' just an appropriate subset of the evolution
equations on the null cone. Again we note that it is not really
the geometry that needs to be averaged, since it is not the
geometry per se that is observed (or appears in the observational
relations). That is, we observe the averaged matter on the null
cone; if we average the geometry both on the null cone, and inside
(and outside) of the null cone, we are averaging quantities that
are not directly observable and may not play any role in
determining the properties of the averaged matter on the null cone
(either directly via the energy momentum tensor or the Ricci
tensor through the EFE, or indirectly through the effects on the
motion of matter in response to the geometry). Thus we would like
to be able to focus on the effects of the averaged geometry that
only affect cosmological observations.

Suppose that it is possible to covariantly `average' just a subset
of the evolution equations on the null cone, focussing on relevant
quantities (and not necessarily the complete averaged spacetime
geometry). Perhaps we can focus just on scalar equations (and not
the full set of EFE) in the spirit of the Buchert approach. In
this approach to the averaging problem (for irrotational dust)
only scalar quantities are averaged, yielding the averaged
Hamiltonian constraint (or generalized Friedmann equation), the
averaged Raychaudhuri equation (plus an integrability condition)
\cite{buch}. The Buchert approach is `heuristic' since the Buchert
equations are not closed. Indeed, since only scalar quantities are
averaged not all of the EFE have been averaged in Buchert's
approach, and consequently any solutions for these equations may
not be consistent with the full set of EFE.

\section{Scalar Equations}

Let $k^a$ represent the tangent vector of the bundle of affinely
parametrized null geodesics (optical rays) with scalar expansion
($\widehat{\theta} = \frac{1}{2} k^a_{~;a}$) and shear
$(\widehat{\sigma})$ \cite{Dyer}.  The null version of the
Raychaudhuri equation is then \cite{Kramer}
\begin{equation}
   D \widehat{\theta} = -\widehat{\theta}^2
- \widehat{\sigma}^2 -\frac{1}{2} R_{ab} {k}^a {k}^b,
\label{Equation1}
\end{equation}
where $D \widehat{\theta}= \widehat{\theta}_{,a} k^a$ is a
directional derivative along the null geodesic (and can be
expressed in terms of the affine parameter, $v$, or the redshift)
and $\widehat{\sigma}^2$ is a scalar. In the Newman-Penrose
formalism, this is effectively eqn. (7.21a) in \cite{Kramer} and
constitutes the first of the optical scalar equations (there is
also an equation for $D(\widehat{\sigma}^2)$).

Let us assume that the matter is dust:
\begin{equation}
T_{ab} = \rho u_a u_b.  \label{Equation2}
\end{equation}
We can decompose the fluid four-velocity as $u^a = k^a +v^a$,
where $V \equiv k^a v_a = k^a u_a ~(v^a v_a = -1 -2V)$ and is
related to the redshift by $1+z = V/V_0$. We can then rewrite eqn.
(\ref{Equation1}) as:
\begin{equation}
D\widehat{\theta} = - \widehat{\theta}^2 - \widehat{\sigma}^2
-\frac{1}{2} \rho V^2.  \label{Equation3}
\end{equation}

The projected part of the conservation equation, $T_{ab}\, ^{;b} =
0$, becomes:
\begin{equation}
D\rho = -\rho\widehat{\theta} - A,         \label{Equation4}
\end{equation}
where $A \equiv \rho_{,b} v^b + \rho v_b\, ^{;b}$.  We can write
an ``effective'' first integral in the form
\begin{equation}
\frac{1}{2} \widehat{\theta}^2 = \rho V^2 + B,
\label{Equation5}
\end{equation}
 where $D(B)$ is obtained from eqns. (\ref{Equation3}) and (\ref{Equation4}).

Equation (\ref{Equation3}) concerns the propagation
 of $\widehat{\theta}$ down the null cone.  Let us average this scalar equation along
 the null cone:
 \begin{equation}
 \frac{d}{d v} \langle \widehat{\theta}\rangle  = -
 \langle \widehat{\theta}\rangle^2 - \frac{V_0^2}{2} \langle \rho \rangle
 (1+z)^2 + \widehat{\cal Q},  \label{Equation7}
 \end{equation}
  where $\widehat{\cal Q} \equiv \langle \widehat{\theta}\rangle^2 - \langle
  \widehat{\theta}^2 \rangle - \langle \widehat{\sigma}^2\rangle$.  We can
  write an averaged
  version of (\ref{Equation5}) in the form
\begin{equation}
 \frac{1}{2} \langle \widehat{\theta}\rangle^2 = V^2_0
 \langle \rho \rangle (1+z)^2
 - \frac{1}{2} \widehat{R} - \frac{1}{2} \widehat{\cal Q}, \label{Equation8}
 \end{equation}
where $\widehat{R}$ is defined through this equation.

Equation (\ref{Equation7}) is an important result; it is a scalar
equation which formally represents how the averaged expansion
varies along the null cone.  It (together with an equation like
(\ref{Equation8})) represents a null version of the Buchert
equations. It suffers the same disadvantages as the Buchert
equations in that not all of the appropriate EFE have been
averaged and the system is not closed.  However, it does have the
advantage of relating physically observed qualities (such as the
density averaged on the null cone), and any assumptions to close
the system now relate correlations on the null cone that may be
physically better motivated.

Alternatively, we can define a characteristic length scale $\widehat{\ell}$ of the average area
 behaviour of the geodesics by:
 $$ \frac{1}{2} \widehat{\theta} = \widehat{\ell}^{-1}
 \frac{d \widehat{\ell}}{d v}.  $$
Eqn.  (\ref{Equation3}) can then be written in the form
  $$  \frac{d^2 \widehat{\ell}}{d v^2} = -2 \widehat{\theta}^2
 - \rho V^2.  $$

\subsubsection{FLRW models}
The spatially homogeneous and isotropic FLRW  metric can be
written in the form:
\begin{equation}
ds^2 = - dt^2 + a^2(t) \left[ dr^2/(1-kr^2) +r^2 d\theta^2 +r^2
\sin^2 \theta d\phi^2 \right]  , \label{eq:metric}
\end{equation}
where $a(t)$ is the cosmic scale factor and $k$ is the curvature
of 3-dimensional space (e.g., $k = 0$ corresponds to a spatially
flat universe). The wavelength $\lambda$ of photons moving through
the universe scale with $a(t)$, and the redshift of light emitted
from a distant source at time $t_{\rm em}$ is defined by $1 + z =
\lambda_{\rm obs}/\lambda_{\rm em} =1/a(t_{\rm em})$; thus $dt =
-dz/H(z)(1+z)$, where $H\equiv \dot a/a$ is the Hubble parameter
and an overdot denotes a time derivative. For an object of
intrinsic luminosity $L$, the measured energy flux $F$ defines the
luminosity distance $d_L$ to the object, where $d_L(z) \equiv
\sqrt{L/4 \pi F} = (1+z)r(z)~,$ and $r(z)$ is the comoving
distance to an object at redshift $z$. The key equations of
cosmology are the Friedmann and Raychaudhuri equations:
\begin{eqnarray}
H^2 = \left(\frac{\dot a}{a}\right)^2 & = & {1 \over 3}\rho
-{k\over a^2} \ \  + {\Lambda\over 3} \label{eq:feq}
\\
\frac{\ddot a}{a}               & = & -{1 \over 6}\,(\rho + 3p) \
\
 + {\Lambda\over 3}
\label{eq:feq2}
\end{eqnarray}
where $\rho$ is the total energy density of the universe, $p$ is
the total pressure, and $\Lambda$ is the cosmological constant.
For each matter component $i$, the separate conservation of energy
is expressed by $d(a^3\rho_i ) = -p_i da^3$.  The deceleration
parameter, $q(z)$, is defined as $q (z) \equiv -{\ddot a\over a
H^2}$.

As an illustration, let us consider equation (\ref{Equation3}) in
the case of a pressure-free, flat $(k=0)$ FLRW model
 with metric (\ref{eq:metric}), where $1 + z = 1/a$. For
 \begin{equation}
 k_a = \left(\frac{1}{a}, 1, 0, 0\right), \label{Equation6}
 \end{equation}
we have by direct calculation  that
\begin{equation}
\widehat{\theta} \equiv \frac{1}{2} k_{a;b}g^{ab} = - \frac{1}{a^2} (\dot{a}
 - \frac{1}{r}),
\end{equation}
 and
\begin{equation}
\widehat{\sigma}^2 \equiv \frac{1}{2} k_{(a;b)} k^{(a;b)}
 - \widehat{\theta}^2 =0,
 \end{equation}
whence equation (\ref{Equation2})  yields
\begin{equation}
a \ddot{a} - \dot{a}^2 = -\frac{1}{2}\rho a^2,
\end{equation}
as expected (and corresponding to eqns. (\ref{eq:feq} -
\ref{eq:feq2})).

{\em Acknowledgements}. I would like to thank Bill Stoeger and
David Wiltshire for helpful comments. This work was supported by
NSERC of Canada.


\begin{thebibliography}{abc}

\baselineskip 12pt


\bibitem{Riess:2006fw} P. Astier {\it et al}, Astron. Astrophys. {\bf 447}, 31
(2006); T.M. Davis {\it et al}, Ap. J. {\bf 666}, 716 (2007).




\bibitem{bennett} J. Dunkley {\em et al}, Ap. J. Suppl. {\bf 180},
306 (2009) ;
D. N. Spergel {\it et al}, Ap. J. Suppl. {\bf 170},
377 (2007).



\bibitem{Kash} A. Kashlinsky {\em et al.}, arXiv:0809.3733; R.
Watkins {\em et al.}, arXiv:0809.4041.

\bibitem{probs3}M. L. McClure and
C. C. Dyer, New. Astron. {\bf 12} 533 (2007).



\bibitem{probs1} S. Ho {\em et al}, Phys. Rev. D
{\bf 78} 043519 (2008).


\bibitem{probs2} C. L. Reichardt {\em et al}, arXiv:0801.1491; U. Seljak {\em et
al}, JCAP {\bf 0610} 014 (2006).

\bibitem{probs4} H. Eriksen {\em et al.}, Ap. J. {\bf 660} L81 (2007);
V. G. Gurzadyan {\em et al.}, arXiv:0807.3652.



\bibitem{Shapiro:2005} C. Shapiro and M. S. Turner, Ap. J. {\bf 649} 563 (2006).



\bibitem{marra-sc} V.~Marra {\em et al},
Phys.\ Rev.\  D {\bf 76}, 123004 (2007); V.~Marra, E.~W.~Kolb and
S.~Matarrese [arXiv:0710.5505].



\bibitem{LTBgeo} H. Alnes {\em et al}  Phys. Rev. D. {\bf 73}
083519 (2006); K. Enqvist and T. Mattsson, JCAP {\bf 0702} 19
(2007); J. Garcia-Bellido and T. Haugbolle, JCAP {\bf 0804} 003
(2008).



\bibitem{CPZ} R. M. Zalaletdinov, Gen. Rel. Grav. {\bf 24} 1015 (1992) \& {\bf
25} 673 (1993); A. A. Coley {\em et al}, Phys. Rev. Letts. {\bf
595} 115102  (2005) [gr-qc/0504115]; A. A. Coley and N. Pelavas,
Phys. Rev. D {\bf 75} 043506 (2006) \& {\bf 74} 087301 (2006).

\bibitem{NewRas} S. R\"{a}s\"{a}nen, JCAP {\bf 0611} 003 (2006) \& JCAP {\bf 0804} 026 (2008)


\bibitem{Col} A. Coley, arXiv:0704.1734.



\bibitem{dunkley}
J. Dunkley {\em et al}, Phys. Rev. Lett. {\bf 95} 261303 (2005).



\bibitem{buch} T. Buchert. Gen. Rel. Grav., {\bf 32} 105 (2000)
 \& {\bf 33} 1381 (2001).


\bibitem{hellaby:volumematching}
C. Hellaby, Gen. Rel. Grav. {\bf 20}, 1203 (1988).


\bibitem{Hoyle:2003} F. Hoyle and M. S. Vogeley ,
Ap. J. {\bf 607} 751 (2004).


\bibitem{Dyer} R.K. Sachs, Proc. Roy. Soc. (London)  {\bf A264} 309
(1974);  C.C. Dyer and R. C. Roeder, Ap. J. {\bf 189} 167 (1974)
\& {\bf180} L31; 167 (1974).


\bibitem{coleynull} A. Coley, arXiv:0812.4565.




\bibitem{R2009}  S. R\"{a}s\"{a}nen, JCAP {\bf 0902} 011 (2009).


\bibitem{Ellis et al} G.F.R. Ellis, S.D. Nel, R. Maartens, W.R. Stoeger,
and A.P. Whitman,
 \textit{Phys. Rep. {\bf 124}},  315 (1985); W.R. Stoeger, G.F.R. Ellis, and S.D. Nel,
 \textit{Class. Quantum Grav. {\bf 9}}, 509 (1992).




\bibitem{AS}
{M.E. Ara\'{u}jo}, {W.R. Stoeger}, {R.C. Arcuri}, {M.L. Bedran},
arXiv:0807.4193.



\bibitem{Kramer}  H. Stephani, D. Kramer, M. A. H. MacCallum, C. A. Hoenselaers and E. Herlt
\textit{Exact solutions of Einstein's field equations} (2nd Ed.
Cambridge University Press, Cambridge, 2003)


\end{thebibliography}
\end{document}